\documentclass[,final ]{aipproc}
\layoutstyle{6x9}

\usepackage{graphicx}
\usepackage{epstopdf}
\usepackage{natbib}
\begin{document}

\title{HI Structure Observations of Reionization and Dark Energy}


\author{Miguel F. Morales}{
  altaddress={Murchison Widefield Array Collaboration},address={University of Washington, Seattle}, altaddress={MIT Kavli Institute},
}

\begin{abstract}
\emph{This proceeding concentrates on the BAO signature of dark energy, and how the SKA dark energy case has been complicated by the emergence of HI structure experiments modeled after the Epoch of Reionization observatories.}
The purpose of the conference talk was to review the current status of the Murchison Widefield Array (MWA), and show the applications of HI structure observations for both reionization and dark energy measurements. Since the status of the MWA is changing weekly, please see the website \cite{MWACollaboration:2008p683} for the current status. This proceedings will instead concentrate on HI structure observations, their applicability to reionization and cosmography, and the implications for the SKA and future HI structure observations of dark energy. 
\end{abstract}

\maketitle


\section{Introduction}
In the last few years efforts to observe the Epoch of Reionization (EoR) via highly redshifted HI emission has become one of the new frontiers of observational cosmology. Several efforts are currently under construction and/or development, including LOFAR, GMRT 150 MHz band, PAPER, EDGES, the MWA and others \cite{Zaroubi:2005p585,Roshi:2006p464,Backer:2007p707,Bowman:2008p180,MWACollaboration:2008p683}. 

While there are a number of different EoR signatures that can be pursued, the majority of the efforts are focused on detecting the three-dimensional power spectrum of HI emission \cite{Zaldarriaga:2004p227,Morales:2004p809,Furlanetto:2006p341}. As recently pointed out by \citet{Wyithe:2008p1710}, approximately 1\% of the hydrogen remains neutral after reionization allowing the measurement of cosmological structure after the end of reionization. (The volume average neutral fraction important for Lyman-$\alpha$ studies is much lower than the 1\% mass average important for 21 cm observations, because the neutral hydrogen is in relatively high density regions.) The natural application of HI structure observations across cosmic time is to the dark energy problem through the measurement of baryon acoustic oscillations (BAO) \cite{Wyithe:2007p2,Chang:2008p439}.

This has particular resonance for the SKA. HI structure experiments such as the MWA do not resolve individual galaxies---they are only concerned about structure at very large scales. However, by concentrating only on large scale structure and survey speed, HI structure experiments could explore the dark energy BAO signature for a small fraction of the SKA's cost. Of course these HI structure observations would miss much of the SKA's science case due to their limited angular resolution, but whether HI dark energy observations are best performed by small dedicated experiments or a full SKA becomes an interesting science question.

In this proceeding, we will briefly review the physics of HI structure observations both during the Epoch of Reionization and afterwards in \S1. We will then look at the sensitivity of HI structure observations and the relevant instrumental scaling relations in \S2, before looking at the implications for the future in \S3.

\section{Fundamentals of HI Power Spectrum Measurements}
\label{HISObsSec}

To date the literature of HI power spectrum observations has focused on the Epoch of Reionization signal. However, the same techniques can be applied to HI observations at all redshifts by just varying the expected neutral hydrogen fraction from $\sim$1 to $\sim$0.01. Often, the best way to study the fundamentals of dark energy HI structure observations is to read the EoR literature, and mentally replace the expected neutral hydrogen and sky temperature terms. In particular, the excellent EoR review paper by \citet{Furlanetto:2006p341} is a good place to start, especially the discussion of observational effects in chapter 9. In this section I will just touch upon a few of the key features of HI structure observations.

Because the HI signal is a redshifted emission line, the observed frequency is a direct measure of the redshift and an indirect measure of the line-of-sight distance. Conceptually, one can envision an HI structure observation as measuring the HI emission with in a large 3D volume of space, with angular position providing the transverse position and observed frequency the line-of-sight position.

At EoR redshifts the velocity distortions are small, and can usually be safely ignored (see \cite{Barkana:2005p595,McQuinn:2006p222} for discussions on extracting fundamental physics from these small velocity distortions). This leads to an approximate 3D symmetry in the observed volume, due to the rotational invariance (isotropy) of space. In the 3D Fourier transform of the observed volume this symmetry implies that the same power is expected within each spherical shell (fixed wavenumber) \cite{Morales:2004p803}. Thus we may envision measuring the HI power spectrum by summing all of the measurements in each spherical annulus to create a 1D measurement of the HI structure, as depicted graphically in Figure \ref{distSensitivity}.

This is very similar to interferometric observations of the cosmic microwave background (CMB), except that the CMB signal is fundamentally a two dimensional surface. In addition to providing many more modes \cite{Loeb:2004p238}, the third dimension offers unique opportunities for subtracting foreground contamination. Most sources of foreground contamination are spectrally smooth---synchrotron and free-free emission. Thus while there is a lot of foreground structure in the plane of the sky, the foregrounds are very smooth in the frequency/line-of-sight direction. This should allow small fluctuations from the redshifted 21 cm emission to be distinguished from the smooth foreground emission \cite{Morales:2006p147,McQuinn:2006p222,Mao:2008p196}. In addition there are spectrally varying foregrounds, such as radio recombination lines, terrestrial interference, beating between Faraday-rotated galactic synchrotron emission and polarized instrumental response, etc. These foregrounds should also be subtractable, as discussed in \citet{Furlanetto:2006p341} and \citet{Morales:2006p147} and references therein.

\section{Sensitivity of HI Structure Observations}
\label{SensitivitySec}

The sensitivity of HI structure observations is somewhat different than standard radio observations due to the unique characteristics of the HI power spectrum. Again we look to the EoR literature, particularly  \cite{Morales:2005p796,Bowman:2006p163,McQuinn:2006p222}.

HI structure observations are fundamentally large scale surveys, and the sensitivity is proportional to the volumetric survey speed---the volume of space observed in a single pointing times the collecting area. This leads to experiments with very large fields of view ($\sim$30$^{\circ}$ FWHM for the MWA). For the EoR the line-of-sight depth of the survey is typically limited by cosmic evolution instead of the instrumental bandwidth, as the observed volume must come from a single epoch and 8 MHz is roughly a redshift interval of $\frac{1}{2}$ at these redshifts. However, at the redshifts appropriate for dark energy the same redshift interval is several hundred MHz and the bandwidth of the observation may become the limiting factor. 

\begin{figure}
\includegraphics{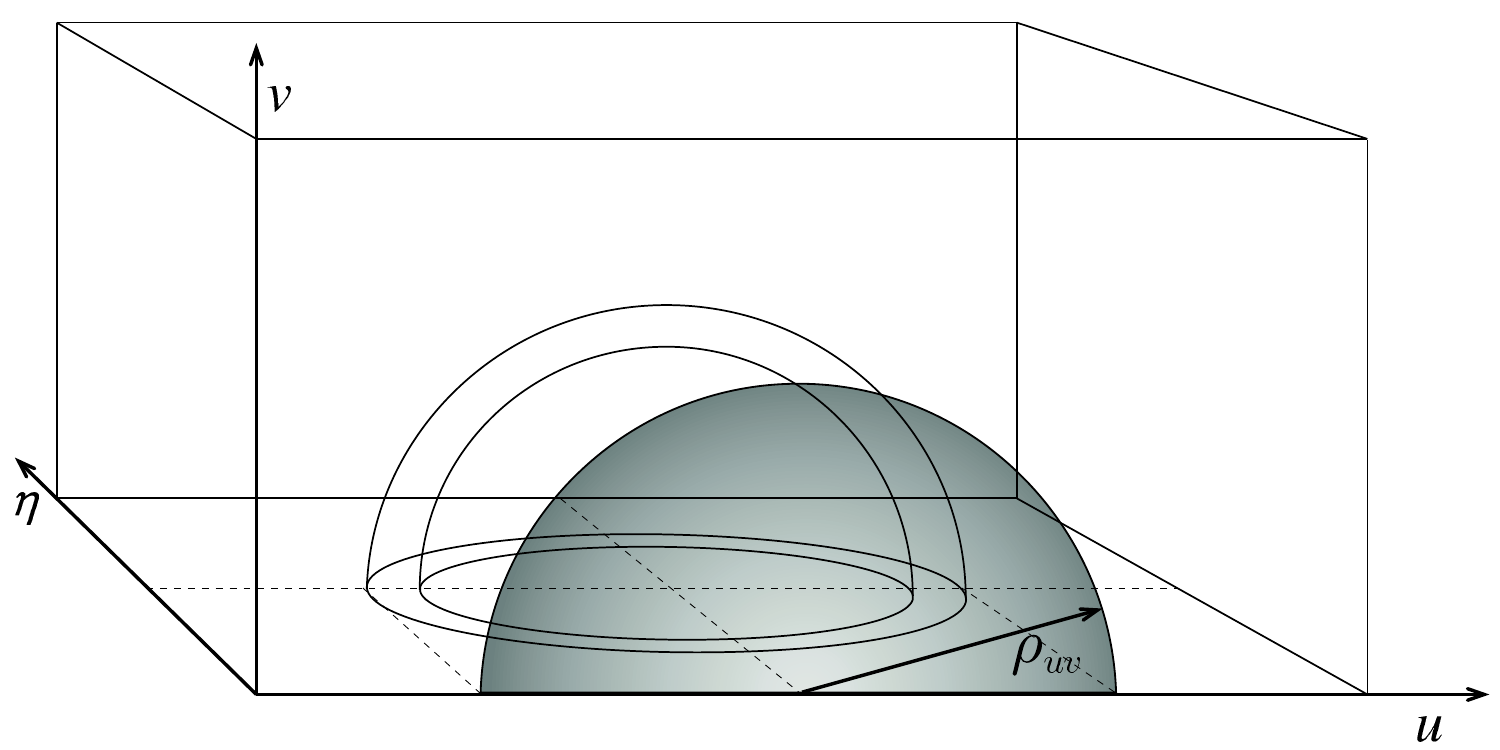}
\caption{The Fourier transform of the observed volume, with $u,v$ representing the spatial wavenumbers and $\eta$ the line-of-sight wavenumber. The measurement of the 21 cm power spectrum at a scale $k$ is performed by combining all of the measurements within a spherical annulus at that diameter, as shown above (the origin is at the center of the annulus where the dashed lines cross).  The shaded region shows the contribtuion of antenna baselines in the $u,v$ plane to this measurement, with dark indicating a higher contribution.  All baselines $\rho_{uv} \leq k$ contribute to the measurement with a limb brightening functional distribution. Even very short baselines contribute information at high $k$ because of the spatial information encoded in the frequency dimension. (Excerpted from \cite{Morales:2005p796}.)}
\label{distSensitivity}
\end{figure}

In addition, HI structure observations benefit from very compact array configurations. Detecting the large scale structure of baryon acoustic oscillations and EoR bubbles is helped by the increased surface brightness sensitivity of a compact array. But somewhat counter-intuitively, the measurement of fine scale structure is also helped by a compact array. This is because even very short baselines can detect small scale structure along the line-of-sight/frequency direction. Thus \emph{all} baselines shorter than the scale of interest contribute to the measurement, as shown in Figure \ref{distSensitivity}. For simple configurations the power spectrum sensitivity is approximaty proportional to the array diameter to the minus fourth power (!). The longer baselines of HI structure experiments are primarily used for calibration and foreground subtraction, as the sensitivity is provided by the short baselines. 

For real array configurations a number of additional factors can come into play, such as the transition from low to high signal-to-noise statistics \cite{Halverson:2002}. Some of the important scaling relationships for HI structure observations are shown in Table  \ref{scalingTable}, and discussed in more length in \citet{Morales:2005p796}.

\begin{table}
\begin{tabular}{rcccccc}
\hline
 &
  \tablehead{1}{c}{b}{$A|_{dA}$} & 
  \tablehead{1}{c}{b}{$A|_{N_{\!A}}$} &
  \tablehead{1}{c}{b}{$dA|_{A}$} &
  \tablehead {1}{c}{b}{ $B$} &
  \tablehead {1}{c}{b}{$k$} &
  \tablehead {1}{c}{b}{$t$} \\
\hline
Power Spec. S/N & $A^{2}$ & $A^{3/2}$ & $(dA)^{-1/2} \propto$  FOV & $B^{1/2}$ &  $k\,\bar{n}(k)$ & $t$  
\end{tabular}
\caption{Approximate scaling relationships for the power spectrum signal-to-noise ratio (excerpted from \cite{Morales:2005p796}).  In order, the scalings in each column are: total array area holding the size of each antenna constant $A|_{dA}$ (adding antennas), total array area holding the number of antennas and distribution constant $A|_{N_{\!A}}$ (increasing antenna size), the size of each antenna with the total array area held constant $dA|_{A}$ (dividing area into more small antennas), the total bandwidth $B$, the sensitivity as a function of wavenumber length $k$ where $\bar{n}(k)$ is the density of baselines (average of $u,v$ coverage as function of wavenumber), and the total observing time $t$.}
\label{scalingTable}
\end{table}

\section{Implications for SKA and Future HI Structure Experiments}
\label{ImplicationsSec}

Because baryon acoustic oscillations occur at very large $\sim$150 Mpc scales, HI structure observations are potentially orders of magnitude more sensitive than the HI galaxy surveys envisioned as part of the SKA science case \cite{Carilli:2004p1169,Carilli:2004p401,Abdalla:2004p1390}. This difference in sensitivity is primarily due to not needing to actually resolve the individual galaxies---for BAO we really only need to know the concentration of galaxies (HI) on $\sim$150 Mpc scales, not the position of any individual galaxy. 

Thus we can imagine an observatory dedicated to measuring the HI BAO that looks very much like a high frequency EoR experiment \cite{Wyithe:2007p2}. It would have a $1/10^{{\rm th}}$ or less of the SKA collecting area, but equivalent or better sensitivity to the dark energy BAO signal.

However, this is not to say that an HI structure experiment is the best way to measure dark energy. First, it is not clear that the BAO signature is the preferred measurement. As highlighted by the Dark Energy Task Force \cite{Albrecht:2006p1021}, systematic errors are often the limiting factor, which could make various weak lensing or galaxy count methods preferable  \cite{Rawlings:2004p1676,Morales:2006p155,Zhang:2005p1612}. Second, an HI structure experiment has a much narrower science focus than the full SKA observatory. While much cheaper than the SKA, if the remainder of the SKA science case motivates the construction of the observatory then the BAO science will come along for ``free.''

What is clear however, is that motivating the SKA via dark energy BAO observations is less compelling than originally thought. The BAO signal is approachable using focused HI structure experiments, at a fraction of the cost.


\begin{theacknowledgments}
I would like to thank Jacqueline Hewitt, Judd Bowman, Avi Loeb and Stuart Wyithe for stimulating discussions. This work has been supported by National Science Foundation grant AST-0457585.
\end{theacknowledgments}

\bibliographystyle{aipproc}
\bibliography{AreciboBib}

\end{document}